\def\be{\begin{equation}}
\def\ee{\end{equation}}
\def\ba{\begin{array}}
\def\ea{\end{array}}

\def\Nb{{I\!\! N}}

\def\Cb{{\Bbb C}}

\documentstyle[12pt]{article}
\input amssym.def
\begin{document}
\parskip=4pt
\parindent=18pt
\baselineskip=22pt \setcounter{page}{1}
\centerline{\Large\bf A Note on Quantum Entanglement and PPT}
\vspace{6ex}
\begin{center}
Shao-Ming Fei$^{1,2}$~~~ and ~~~Xianqing Li-Jost$^3$
\bigskip

\begin{minipage}{6.2in}
$^1$Department of Mathematics, Capital Normal University, Beijing 100037, P.R. China\\
$^2$Institut f{\"u}r Angewandte Mathematik, Universit{\"a}t Bonn, 53115 Bonn, Germany\\
E-mail: fei@iam.uni-bonn.de\\
$^3$Max-Planck-Institute for Mathematics in the Sciences, 04103 Leipzig, Germany\\
E-mail: Xianqing.Li-Jost@mis.mpg.de

\end{minipage}
\end{center}
\vskip 1 true cm
\parindent=18pt
\parskip=6pt
\begin{center}
\begin{minipage}{5in}
\vspace{3ex} \centerline{\large Abstract} \vspace{4ex}

We study quantum states for which the PPT criterion is both sufficient and necessary
for separability. We present a class of $3\times 3$ bipartite mixed states
and show that these states are separable if and only if they are PPT.
\end{minipage}
\end{center}


\vfill
\smallskip
\bigskip

\section{Introduction}

Quantum entanglement has been
recently the subject of much study as a potential resource for communication and
information processing \cite{nielsen}.
Thus characterization and quantification of entanglement become an important issue.
Entanglement of formation (EOF)
\cite{eof} and concurrence \cite{concurrence} are two well
defined quantitative measures of quantum entanglement. For two-quibt
systems it has been proved that EOF is a monotonically increasing
function of the concurrence and an elegant formula for the
concurrence was derived analytically by Wootters \cite{wotters}.
However with the increasing dimensions of the subsystems the
computation of EOF and concurrence becomes formidably difficult. A
few explicit analytic formulae for EOF and concurrence have been
found only for some special symmetric states
\cite{Terhal-Voll2000}.

In fact if one only wants to know wether a state is separable or not,
it is not necessary to compute the exact values of the measures for quantum entanglement.
The estimation of lower
bounds of entanglement measures can be just used to judge the separability
of a quantum state \cite{chenk}. There are also many separability criteria, e.g.,
PPT (positive partial transposition) criterion
\cite{peres}, realignment \cite{ChenQIC03} and generalized realignment criteria
\cite{chenkai}, as well as some necessary and sufficient operational
criteria for low rank density matrices \cite{hlvc00}.
Further more, separability criteria based on local uncertainty
relation \cite{hofmann} and the correlation
matrix \cite{julio} of the Bloch representation for a quantum
state have been derived, which are strictly stronger than or
independent of the PPT and realignment criteria.

The PPT criterion is generally a necessary condition for separability. It becomes
sufficient only for the cases $2\times2$ and
$2\times3$ bipartite systems \cite{2231}. Other states of such property
are the Schmidt-correlated (SC) states \cite{SC}, which are the mixtures of pure states
sharing the same Schmidt bases and naturally appear in a bipartite system
dynamics with additive integrals of motion \cite{SC1}. In this paper we consider
another special class of $3\times 3$ quantum mixed states. We show that for
the states in this class, PPT is both necessary and sufficient for separability.

\section{A class of quantum states and PPT}

We consider $3\times 3$ quantum mixed states given by
\be\label{rho}
\rho = \lambda |X><X| + \lambda' |X'><X'| + \lambda'' |X''><X''|,
\ee
where $\lambda  + \lambda' + \lambda''=1$, $0<\lambda , \lambda',
\lambda''<1$, $|X>, |X'>,|X''>$ are orthonormal vectors,
\be\label{XXX}
\ba{l}
|X> = (\alpha,0,0,0,\beta,0,0,0,\gamma)^t,\\[2mm]
|X'>= (0,\alpha',0,0,0,\beta',\gamma',0,0)^t,\\[2mm]
|X''>= (0,0,\alpha'',\beta'',0,0,0,\gamma'',0)^t,
\ea
\ee
where $\alpha,\beta,\gamma,\alpha',\beta',\gamma',\alpha'',\beta'',\gamma''$
are non-zero complex numbers,
$t$ stands for transposition. If we take the basis $|1>=(1,0,0)^t$,
$|2>=(0,1,0)^t$, $|3>=(0,0,1)^t$, then
$|X(\alpha,\beta,\gamma)>\equiv |X>
=\alpha|11>+\beta|22>+\gamma|33>$, $|X'> =(I\otimes
P)|X(\alpha',\beta',\gamma')>$, $|X''> =(I\otimes
P^2)|X(\alpha'',\beta'',\gamma'')>$, where
$P=\left(\ba{ccc}0&0&1\\1&0&0\\0&1&0\ea\right)$ is the permutation
operator.

{\sf Theorem:} State $\rho$ is separable if and only if it is PPT.

To prove the theorem we first note that after partial transposition
$\rho$ has the form
$$\ba{l}
\rho^{pt}=\\[2mm]
\left(
\begin{array}{ccccccccc}
 \lambda\alpha\alpha^*&0&0&0&0&\lambda''\alpha''\beta''^*&0&\lambda'\alpha'\gamma'^*&0\\[3mm]
 0&\lambda'\alpha'\alpha'^*&0&\lambda\alpha\beta^*&0&0&0&0&\lambda''\alpha''\gamma''^*\\[3mm]
 0&0&\lambda''\alpha''\alpha''^*&0&\lambda'\alpha'\beta'^*&0&\lambda\alpha\gamma^*&0&0\\[3mm]
 0&\lambda\alpha^*\beta&0&\lambda''\beta''\beta''^*&0&0&0&0&\lambda'\beta'\gamma'^*\\[3mm]
 0&0&\lambda'\alpha'^*\beta'&0&\lambda\beta\beta^*&0&\lambda''\beta''\gamma''^*&0&0\\[3mm]
 \lambda''\alpha''^*\beta''&0&0&0&0&\lambda'\beta'\beta'^*&0&\lambda\beta\gamma^*&0\\[3mm]
 0&0&\lambda\alpha^*\gamma&0&\lambda''\beta''^*\gamma''&0&\lambda'\gamma'\gamma'^*&0&0\\[3mm]
 \lambda'\alpha'^*\gamma'&0&0&0&0&\lambda\beta^*\gamma&0&\lambda''\gamma''\gamma''^*&0\\[3mm]
 0&\lambda''\alpha''^*\gamma''&0&\lambda'\beta'^*\gamma'&0&0&0&0&\lambda\gamma\gamma^*
\end{array}
\right).
\ea
$$
$\rho^{pt}$ is hermitian. The non-negativity of $\rho^{pt}$, $\rho^{pt}\geq 0$,
implies that $<\psi|\rho^{pt}|\psi> \geq 0$ for all vector $|\psi>\in H\otimes H$, which is equivalent
to the non-negativity of the following three $3\times 3$ matrices:
\be A_1=\left(
\begin{array}{ccc}
 \lambda\alpha\alpha^*&\lambda''\alpha''\beta''^*&\lambda'\alpha'\gamma'^*\\[3mm]
 \lambda''\alpha''^*\beta''&\lambda'\beta'\beta'^*&\lambda\beta\gamma^*\\[3mm]
 \lambda'\alpha'^*\gamma'&\lambda\beta^*\gamma&\lambda''\gamma''\gamma''^*
\end{array}\right),
\ee
\be A_2=\left(
\begin{array}{ccc}
\lambda'\alpha'\alpha'^*&\lambda\alpha\beta^*&\lambda''\alpha''\gamma''^*\\[3mm]
\lambda\alpha^*\beta&\lambda''\beta''\beta''^*&\lambda'\beta'\gamma'^*\\[3mm]
\lambda''\alpha''^*\gamma''&\lambda'\beta'^*\gamma'&\lambda\gamma\gamma^*
\end{array}\right),
\ee
and
\be A_3=\left(
\begin{array}{ccc}
\lambda''\alpha''\alpha''^*&\lambda'\alpha'\beta'^*&\lambda\alpha\gamma^*\\[3mm]
\lambda'\alpha'^*\beta'&\lambda\beta\beta^*&\lambda''\beta''\gamma''^*\\[3mm]
\lambda\alpha^*\gamma&\lambda''\beta''^*\gamma''&\lambda'\gamma'\gamma'^*
\end{array}\right).
\ee

The non-negativity of $A_1$ is equivalent to the following inequalities:
\begin{eqnarray}\label{rcl}\label{11}
\lambda\lambda'|\alpha|^2|\beta'|^2&\geq&\lambda''^2|\alpha''|^2|\beta''|^2,\\\label{12}
\lambda\lambda''|\alpha|^2|\gamma'''|^2&\geq&\lambda'^2|\alpha'|^2|\gamma'|^2,\\\label{13}
\lambda'\lambda''|\beta'|^2|\gamma''|^2&\geq&\lambda^2|\gamma|^2|\beta|^2
\end{eqnarray}
and
\be\label{14}
\ba{l}
\lambda\lambda'\lambda''|\alpha|^2|\beta'|^2|\gamma''|^2+2\lambda\lambda'
\lambda''Re\alpha'\alpha''^*\beta''\beta^*\gamma\gamma'^*
-\lambda^3|\alpha|^2|\beta|^2|\gamma|^2\\[2mm]
-\lambda'^3|\alpha'|^2|\beta'|^2|
\gamma'|^2-\lambda''^3|\alpha''|^2|\beta''|^2|\gamma''|^2\geq 0.
\ea
\ee
Similarly the non-negativity of $A_2$ and $A_3$ give rise to
\begin{eqnarray}\label{21}
\lambda'\lambda''|\alpha'|^2|\beta''|^2\geq\lambda^2|\alpha|^2|\beta|^2,\\ \label{22}
\lambda\lambda'|\alpha'|^2|\gamma|^2\geq\lambda''^2|\alpha''|^2|\gamma''|^2,\\\label{23}
\lambda\lambda''|\gamma|^2|\beta''|^2\geq\lambda'^2|\gamma'|^2|\beta'|^2,
\end{eqnarray}
\be\label{24}
\ba{l}
\lambda\lambda'\lambda''|\alpha'|^2|\beta''|^2|\gamma|^2+2\lambda\lambda'
\lambda''Re\alpha\alpha''^*\beta'\beta^*\gamma''\gamma'^*
-\lambda^3|\alpha|^2|\beta|^2|\gamma|^2\\[2mm]
-\lambda'^3|\alpha'|^2|\beta'|^2|
\gamma'|^2-\lambda''^3|\alpha''|^2|\beta''|^2|\gamma''|^2
\geq 0,
\ea
\ee
and
\begin{eqnarray}\label{31}
\lambda\lambda''|\alpha''|^2|\beta|^2\geq\lambda'^2|\alpha'|^2||\beta'|^2,\\\label{32}
\lambda'\lambda''|\alpha''|^2|\gamma'|^2\geq\lambda^2|\alpha|^2|\gamma|^2,\\\label{33}
\lambda\lambda'|\beta|^2|\gamma'|^2\geq\lambda''^2|\beta''|^2|\gamma''|^2,
\end{eqnarray}
\be\label{34}
\ba{l}
\lambda\lambda'\lambda''|\alpha''|^2|\beta|^2|\gamma'|^2+2\lambda\lambda'
\lambda''Re\alpha'\alpha^*\beta'^*\beta''\gamma''^*\gamma
-\lambda^3|\alpha|^2|\beta|^2|\gamma|^2\\[2mm]
-\lambda'^3|\alpha'|^2|\beta'|^2|
\gamma'|^2-\lambda''^3|\alpha''|^2|\beta''|^2|\gamma''|^2\geq 0.
\ea
\ee

We can show that the inequalities (\ref{11})-(\ref{13}), (\ref{21})-(\ref{23}),
(\ref{31})-(\ref{33}) are equalities. In fact if (\ref{11}) is an inequality,
$\lambda\lambda'|\alpha|^2|\beta'|^2 > \lambda''^2|\alpha''|^2|\beta''|^2$,
then from (\ref{21}) and (\ref{31}), one would have,
by multiplying $\lambda'|\alpha'|^2$ on both sides,
$\lambda'|\alpha'|^2\lambda''^2|\alpha''|^2|\beta''|^2
<\lambda\lambda'^2|\alpha|^2|\alpha'|^2|\beta'|^2
\le\lambda\lambda''|\alpha''|^2|\beta|^2\lambda|\alpha|^2
\le \lambda'|\alpha'|^2\lambda''^2|\alpha''|^2|\beta''|^2$, which contradicts.
Therefore (\ref{11})-(\ref{13}), (\ref{21})-(\ref{23}),
(\ref{31})-(\ref{33}) become
\begin{eqnarray}\label{41}
\sqrt{\lambda\lambda'}\alpha\beta'=\lambda''\alpha''\beta'' e^{i\theta''_1},~
\sqrt{\lambda\lambda''}\alpha\gamma''=\lambda'\alpha'\gamma' e^{i\theta'_2},~
\sqrt{\lambda''\lambda'}\gamma''\beta'=\lambda\gamma\beta e^{i\theta_3}\\\label{42}
\sqrt{\lambda''\lambda'}\alpha'\beta''=\lambda\alpha\beta e^{i\theta_1},~
\sqrt{\lambda\lambda'}\alpha'\gamma=\lambda''\alpha''\gamma'' e^{i\theta''_2},~
\sqrt{\lambda\lambda''}\gamma\beta''=\lambda'\gamma'\beta' e^{i\theta'_3}\\\label{43}
\sqrt{\lambda\lambda''}\alpha''\beta=\lambda'\alpha'\beta' e^{i\theta'_1},~
\sqrt{\lambda''\lambda'}\alpha''\gamma'=\lambda\alpha\gamma e^{i\theta_2},~
\sqrt{\lambda\lambda'}\gamma'\beta=\lambda''\gamma''\beta'' e^{i\theta''_3},
\end{eqnarray}
where all $\theta\in[-\pi,\pi]$.
From (\ref{41})-(\ref{43}), (\ref{14}), (\ref{24}) and (\ref{34}) become
\begin{eqnarray}
Re(\alpha'\alpha''^*\beta''\beta^*\gamma\gamma'^*)\geq|\alpha|^2|\beta'|^2|\gamma''|^2,\\
Re(\alpha\alpha''^*\beta'\beta^*\gamma''\gamma'^*)\geq|\alpha'|^2|\beta''|^2|\gamma|^2,\\
Re(\alpha'\alpha^*\beta'^*\beta''\gamma''^*\gamma)\ge|\alpha''|^2|\beta|^2|\gamma'|^2.
\end{eqnarray}

Applying these 9 equalities one obtains also
\be\begin{array}{cc}
|\alpha'\alpha''^*\beta''\beta^*\gamma\gamma'^*|&\leq |\alpha'||\alpha''||\beta''||\beta||\gamma||\gamma'|\\[3mm]
=\frac{\lambda^2}{\lambda'\lambda''}|\alpha||\beta||\alpha||\gamma||\beta||\gamma|
&=|\alpha|^2|\beta'|^2|\gamma''|^2.
\end{array}
\ee
Similarly one has
\be
|\alpha\alpha''^*\beta'\beta^*\gamma''\gamma'^*|\le|\alpha'|^2|\beta''|^2|\gamma|^2,~~~
|\alpha'\alpha^*\beta'^*\beta''\gamma''^*\gamma|\le|\alpha''|^2|\beta|^2|\gamma'|^2.
\ee
By using (\ref{41})-(\ref{43}), we get
\begin{eqnarray}\label{51}
Re(\alpha'\alpha''^*\beta''\beta^*\gamma\gamma'^*) = \alpha'\alpha''^*\beta''\beta^*\gamma\gamma'^*=
|\alpha|^2|\beta'|^2|\gamma''|^2,\\\label{52}
Re(\alpha\alpha''^*\beta'\beta^*\gamma''\gamma'^*)=\alpha\alpha''^*\beta'\beta^*\gamma''\gamma'^* =
|\alpha'|^2|\beta''|^2|\gamma|^2,\\\label{53}
Re(\alpha'\alpha^*\beta'^*\beta''\gamma''^*\gamma) =\alpha'\alpha^*\beta'^*\beta''\gamma''^*\gamma =
|\alpha''|^2|\beta|^2|\gamma'|^2.
\end{eqnarray}
From (\ref{51}) and (\ref{52}), we have
$$
\frac{\alpha'\beta''\gamma}{\alpha\beta'\gamma''}=\frac{|\alpha|^2|
 \beta'|^2|\gamma''|^2}{|\alpha'|^2|\beta''|^2|\gamma|^2}
=\frac{\lambda^2|\alpha|^2\cdot\frac{1}{\lambda'\lambda''}|\beta|^2|\gamma|^2}{|\alpha'|^2|
\beta''|^2|\gamma|^2}=1.
$$
While from (\ref{51}) and (\ref{53}), we have
$$
(\frac{\alpha''\beta\gamma'}{\alpha\beta'\gamma''})^*=1.
$$
Therefore
$$
\alpha'\beta''\gamma=\alpha\beta'\gamma''=\alpha''\beta\gamma'
$$
and, from (\ref{41})-(\ref{43}),
$$
\theta_1=\theta_2=\theta_3\equiv\theta ~,\theta'_1=\theta'_2=\theta'_3\equiv\theta'~,
\theta''_1=\theta''_2=\theta''_3\equiv\theta''.
$$

Now by using these PPT conditions of $\rho$ we prove that $\rho$ has a pure
state decomposition $\rho =\sum_{l=1}^{l=3}|\psi_l><\psi_l|$ such that all states
$|\psi_l>$, $l=1,2,3$, are separable. $|\psi_l>$ can be generally expressed as
$|\psi_l>=\sum_m^3 U_{ml}|X_m>=\sum_{ij}^{3} a_{ij}^{l}|ij>$ for some
$a_{ij}^{l}\in\Cb$ under some basis $|ij>$, where
$U_{ml}$ are the entries of a $3\times 3$ unitary matrix $U$,
$|X_1>=\sqrt{\lambda}|X>$, $|X_2>=\sqrt{\lambda'}|X'>$, $|X_3>=\sqrt{\lambda''}|X''>$.
We denote $B_l$ the $3\times 3$ matrix with entries $a_{ij}^{l}$.
Suppose the matrix $U$ has the following form
\be
U= \left( \begin{array}{ccc} u_1&u_2&u_3\\[3mm]
 u'_1 &u'_2&u'_3\\[3mm]
 u''_1 &u''_2&u''_3 \\[3mm]
\end{array} \right)
= \left( \begin{array}{ccc} c_1 e^{i\theta}
 &c_2 e^{i\theta}&c_3 e^{i\theta}\\[3mm]
 c'_1 e^{i\theta'}&c'_2 e^{i\theta'}&c'_3 e^{i\theta'}\\[3mm]
 c''_1 e^{i\theta''}&c''_2 e^{i\theta''}
&c''_3 e^{i\theta''}
\end{array} \right),
\ee
where according to the unitary condition of $U$,
\be\label{ud}
\sum_{l=1}^{l=3} c_lc_l^*=\sum_{l=1}^{l=3}c'_lc'^*_l = \sum_{l=1}^{l=3}c''_lc''^*_l = 1,~
\sum_{l=1}^{l=3}c_lc'^*_l= 0,~\sum_{l=1}^{l=3}c_lc''^*_l = 0, ~\sum_{l=1}^{l=3}c'_lc''^*_l= 0.
\ee
Then $B_l$, $l=1,2,3$, has the following form
 \be
 B_l=\left(
\begin{array}{ccc}
 \sqrt{\lambda}\alpha u_l&\sqrt{\lambda'}\alpha'u'_l&\sqrt{\lambda''}\alpha''u''_l\\[3mm]
 \sqrt{\lambda''}\beta''u''_l&\sqrt{\lambda}\beta u_l&\sqrt{\lambda'}\beta'u'_l\\[3mm]
 \sqrt{\lambda'}\gamma'u'_l&\sqrt{\lambda''}\gamma''u''_l&\sqrt{\lambda}\gamma u_l
\end{array}
\right).
\ee
It is straightforward to verify that the matrix $B_l$ has rank one if
\be\label{rankone}
c_l^2 e^{i2\theta}=c'_lc''_l.
\ee

As $0 < rank(B_l B_l^+) \leq rank (B_l)rank (B_l^+)$, if the rank of $B_l$ is one,
matrix $B_l B_l^+$ has also rank one and $|\psi_l>$ is separable.
Therefore if we can find $c_l$, $c'_l$, $c''_l$, $l = 1,2,3$, satisfying
the unitary condition (\ref{ud}) and the rank one condition (\ref{rankone}),
then $\rho = \lambda |X><X| + \lambda' |X'><X'| + \lambda'' |X''><X''|$
has separable pure state decomposition,
$\rho =\sum_{l=1}^{l=3}|\psi_l><\psi_l|$
and $\rho$ is then separable if it is PPT.
We show now that there exist such coefficients $c_i$, $c'_i$, $c''_i$, $i = 1,2,3$,
satisfying both the unitary condition and the rank-one condition.
Set $c_l = \frac{1}{\sqrt{3}}e^{i\varphi_l}$, $c'_l =
\frac{1}{\sqrt{3}}e^{i\varphi'_l}$, $c''_l =
\frac{1}{\sqrt{3}}e^{i\varphi''_l}$, $l = 1,2,3$, with
$\varphi_l$, $\varphi'_l$, $\varphi''_l$, $l = 1,2,3$, satisfying
$$
\varphi_1-\varphi'_1=\xi',~ \varphi_2-\varphi'_2
=\xi'+ \frac{2\pi}{3},~ \varphi_3-\varphi'_3=\xi'-\frac{2\pi}{3},
$$
$$\varphi_1-\varphi''_1=\xi'' + \frac{2\pi}{3},~ \varphi_2-\varphi''_2=\xi'',~
\varphi_3-\varphi''_3=\xi''-\frac{2\pi}{3},$$
for some real numbers $\xi'$ and $\xi''$.
Then the unitary conditions (\ref{ud}) are satisfied.

The rank-one conditions require that
$-2\varphi_l + \varphi'_l+\varphi''_l= 2\theta,~
-2\varphi'_l + \varphi_l+\varphi''_l= 2\theta',~
-2\varphi''_l + \varphi_l+\varphi'_l= 2\theta''$, which can be
realized by simply choosing
$\xi'=\frac{2}{3}\theta'-\frac{2}{3}\theta$,
$\xi''=-\frac{2}{3}\theta'-\frac{4}{3}\theta- \frac{2\pi}{3}$.
Therefore if state $\rho$ is PPT, then it is separable.
In fact due to that there is still freedom in choosing the
parameters $\varphi_l$, $\varphi'_l$, $\varphi''_l$, $l=1,2,3$,
there exist many separable pure state decompositions.

\section{Conclusions and remarks}

We have studied a special kind of bipartite quantum mixed states.
For this class of states, the PPT criterion is both sufficient and necessary for separability.
Here the states we concerned are rank three ones on $3\times 3$ bipartite systems.
It has been shown that any bipartite entangled states of rank three are distillable \cite{rankthree},
that is, there is no rank three bipartite bound entangled states.
Therefore if the state $\rho$ is not PPT, i.e. conditions
$$
\frac{\lambda}{\sqrt{\lambda'\lambda''}}|\alpha\beta\gamma|
=\frac{\lambda'}{\sqrt{\lambda\lambda''}}|\alpha'\beta'\gamma'|
=\frac{\lambda''}{\sqrt{\lambda\lambda'}}|\alpha''\beta''\gamma''|
$$
are not satisfied, $\rho$ must be not only entangled, but also distillable.
This gives an example that a separable state could directly become a distillable state
when some parameters varies continuously.
There could be no bound entangled states between separable states and
distillable states. Above all, with a similar construction of states (\ref{XXX}),
rank $2k+1$ states on $(2k+1)\times (2k+1)$, $k\in \Nb$, bipartite system can be obtained.
Analogous investigations could be applied to get similar results.

\bigskip
\noindent{\bf Acknowledgments}\,
This work is supported by the NSFC 10675086, NSFC 10875081,
KZ200810028013 and NKBRPC (2004CB318000).

\end{document}